\documentclass[11pt,letterpaper]{article}

\usepackage{jheppub}
\usepackage{braket}
\usepackage{amsmath}
\usepackage{amsfonts}
\usepackage{amssymb}
\usepackage[force]{feynmp-auto}
\usepackage{graphicx}
\usepackage{epsfig}
\usepackage{color}
\usepackage{hyperref}
\usepackage{enumitem}
\usepackage{multirow}
\usepackage{todonotes}

\def\ba{\begin{eqnarray}}
\def\ea{\end{eqnarray}}
\def\be{\begin{equation}}
\def\ee{\end{equation}}
\def\bea{\begin{eqnarray}}
\def\eea{\end{eqnarray}}
\def\bef{\begin{flalign}}
\def\eef{\end{flalign}}
\def\nn{\nonumber}
\def\d{\mathrm{d}}
\def\vk{\vec{k}}
\def\vp{\vec{p}}
\def\vx{\vec{x}}

\def\hPhi{\hat{\Phi}}
\def\la{\langle}
\def\ra{\rangle}
\def\({\left(}
\def\){\right)}
\def\[{\left[}
\def\]{\right]}
\def\<{\left\langle}
\def\>{\right\rangle}

\def\tt{\textcolor{violet}}

\title{\center{Loop corrections in Minkowski spacetime away from equilibrium. Part II. Finite-time results}}

\author[a]{Spasen Chaykov,}
\author[a]{Nishant Agarwal,}
\author[b]{Sina Bahrami,}
\author[c]{and R. Holman}

\affiliation[a]{Department of Physics and Applied Physics, University of Massachusetts, Lowell, MA 01854, USA}
\affiliation[b]{Institute for Gravitation and the Cosmos, The Pennsylvania State University, University Park, PA 16802, USA}
\affiliation[c]{Minerva University, 14 Mint Plaza, Suite 300, San Francisco, CA 94103, USA}

\emailAdd{spasen\_chaykov@student.uml.edu}
\emailAdd{nishant\_agarwal@uml.edu}
\emailAdd{sb933@cornell.edu}
\emailAdd{rh4a@andrew.cmu.edu}

\abstract{Loop corrections to finite-time correlation functions in quantum field theories away from equilibrium can be calculated using the in-in path integral approach. In this paper, we calculate the unequal-time two-point correlator for different massless self-interacting scalar quantum field theories on a Minkowski background, starting the field evolution at an arbitrary initial time. We find the counterterms that need to be added to UV-renormalize the result, including usual in-out counterterms in the dynamics and additional initial state counterterms that are required to cancel all UV divergences. We find that the late-time limit of the renormalized correlation function exhibits a linear or logarithmic growth in time, depending on whether the interaction strength is dimension-one or dimensionless, respectively. The late-time correlations match those obtained in our companion paper and, as shown there, the divergences do not indicate a real IR issue, consistent with what one would expect in Minkowski.}

\keywords{Non-equilibrium field theory, Renormalization and regularization, Effective field theories}
\arxivnumber{2206.11289}

\begin{document}

\maketitle

%%%%%%%%%%%%%%%%%%%%%%%%%%%%%%%%%%%%%%%%%%%%%%%%%%

%------------------------------------------------
\section{Introduction}
\label{sec:intro}
%------------------------------------------------

The dynamics of an out-of-equilibrium quantum system can be described in terms of its correlation functions. The two-point correlation, for example, describes the linear response of a quantum system and is a central object in the study of various out-of-equilibrium phenomena including, for example, inflation in the early Universe, quantum chaos in many-body systems, and thermalization, or lack thereof, in open quantum systems. With this in mind, it is crucial to understand how the two-point correlation changes in the presence of nonlinearities, for example, whether it picks up perturbative corrections or acquires a qualitatively different non-perturbative character.

A standard method to calculate finite-time correlation functions in out-of-equilibrium many-body quantum systems and quantum field theory is the in-in path integral approach \cite{Schwinger:1960qe,Mahanthappa:1962ex,Bakshi:1962dv,Kadanoff:1962,Bakshi:1963bn,Keldysh:1964ud,Jordan:1986ug,Calzetta:1986ey}. In cosmology, it is most importantly used to obtain equal-time correlations of the primordial fluctuation and loop corrections in the presence of gravity-induced interactions. The formalism also allows, of course, the calculation of unequal-time correlations. In this paper, we use it to  calculate loop corrections to the unequal-time two-point correlator of a massless scalar field $\phi(\vx,t)$ in Minkowski spacetime and in Fourier space. We assume that the field is in the vacuum of the free theory until the interaction is switched on instantaneously at the time $t_0$. As shown in the paper, this choice is not always justified since the act of switching on the interaction can itself create excitations in the initial state or, in other words, the system is not in the free theory ground state at any finite time after $t_0$. We, nevertheless, make this assumption to simplify our calculations, and find that it is imperative to include perturbative corrections to the initial state. The specific interactions that we consider are $\lambda\phi^3$ in 4D and 6D and $\lambda\phi^4$ in 4D.

In order to regulate our loop integrals, we introduce a variant of dimensional regularization in which we first convert to a Euclidean time coordinate, which damps out the mode function at high momenta and allows us to easily perform spatial Fourier integrals, and then change the dimension of the time integral in the loop. We found this regularization scheme to be more reliable than a hard cutoff or damping out the mode function via an $i\epsilon$ prescription. We renormalize the resulting correlation function by adding two types of counterterms, the first are usual in-out counterterms in the dynamics that respect the background Lorentz symmetry, and the second are counterterms in the initial state that we alluded to above. The presence of initial state counterterms has been emphasized before \cite{Calzetta:2008,Baacke:1997zz,Baacke:1999ia,Collins:2005nu,Collins:2014qna} and our results further highlight their importance and ubiquity in finite-time calculations.

At second order in perturbation theory, we find that the two-point correlator grows secularly as the difference of the two times increases. The growth is linear for $\lambda\phi^3$ in 4D (where $\lambda$ has dimensions of mass) and logarithmic for both $\lambda\phi^3$ in 6D and $\lambda\phi^4$ in 4D (where $\lambda$ is dimensionless). This agrees with the findings of our companion paper \cite{Chaykov:2022zro}, where we first obtain the late-time result using standard techniques of in-out perturbation theory, assuming that the interaction is switched on adiabatically in the infinite past, and then show that the late-time divergences can in fact be resummed into late-time decays.

The paper is organized as follows. We set up our notation and the problem in section\ \ref{sec:setup}. In section\ \ref{sec:inin}, we briefly review the calculation of correlation functions using the in-in formalism and discuss how to add counterterms in the initial state. We calculate the unequal-time two-point correlator for different interactions, including all renormalization counterterms, in section\ \ref{sec:inincalc}. We end with a discussion in section\ \ref{sec:disc}.

%%%%%%%%%%%%%%%%%%%%%%%%%%%%%%%%%%%%%%%%%%%%%%%%%%

%------------------------------------------------
\section{Setup}
\label{sec:setup}
%------------------------------------------------

The basic setup in this paper is the same as that of our companion paper \cite{Chaykov:2022zro}, except that the results here are valid at any time while \cite{Chaykov:2022zro} focuses on the late-time limit. Consider the Schr\"{o}dinger picture field operator $\hPhi_S(\vx)$ with eigenstates $| \phi(\cdot) \rangle$ and eigenvalues $\phi(\vx)$, so that $\hPhi_S(\vx) | \phi(\cdot) \rangle = \phi(\vx) | \phi(\cdot) \rangle$. The dot indicates all field configurations and the completeness relation is given by $\int {\cal D} \phi(\cdot)| \phi(\cdot) \rangle \langle \phi(\cdot) | = \hat{1}$. We are interested in calculating the spatial Fourier transform of the connected correlation function ${\rm Tr} \big[ \hat{\rho}(t_0) \hPhi(\vx,t) \hPhi(\vx',t') \big]_c$, with $t > t'$, for different self-interacting field theories; here $\hat{\rho}(t_0)$ is the initial density operator for the field, which we choose to coincide with the free theory's ground state, and $\hPhi(\vx,t)$ is the Heisenberg picture field operator. We use the in-in path integral approach to do this calculation at finite times, so that $t$ and $t'$ do not have to be much greater than $t_0$.

We specifically consider the action $S[\phi] = \int \d^dx \, {\cal L}[\phi]$ in $d$-dimensional Minkowski spacetime with the following two Lagrangian densities,
\bea
    {\cal L}_3[\phi] & = & -\frac{1}{2} (\partial_{\alpha}\phi)^2 - \frac{1}{2} m^2 \phi^2 - \frac{1}{3!} \lambda \phi^3 - \frac{1}{2} \delta_r (\partial_{\alpha}\phi)^2 - \frac{1}{2} \delta_m \phi^2 - \frac{1}{3!} \delta_\lambda \phi^3 + Y\phi \quad
\label{eq:lag3}
\eea
and
\bea
    {\cal L}_4[\phi] & = & -\frac{1}{2} (\partial_{\alpha}\phi)^2 - \frac{1}{2} m^2 \phi^2 - \frac{1}{4!} \lambda \phi^4 - \frac{1}{2} \delta_r (\partial_{\alpha}\phi)^2 - \frac{1}{2} \delta_m \phi^2 - \frac{1}{4!} \delta_\lambda \phi^4 \, ,
\label{eq:lag4}
\eea
where $(\partial_{\alpha}\phi)^2 = -\dot{\phi}^2 + (\partial_i\phi)^2$ with the dot now indicating a derivative with time, $m$ is the mass parameter, $\lambda$ is a coupling constant, and the terms with $\delta_r$, $\delta_m$, $\delta_{\lambda}$, and $Y$ are the usual counterterms required to cancel any UV divergences. As discussed in the next section, additional initial state counterterms are required to fully cancel the UV divergences that appear in finite-time correlation functions. 

We treat the $\lambda$ terms perturbatively, restricting our calculations to second order in perturbation theory, or ${\cal O}(\lambda^2)$. Further, we restrict our calculations to the massless ($m = 0$) case for technical reasons. At this order and with $m$ set to zero, specific counterterms may or may not contribute, in particular, $\delta_{\lambda}$ does not contribute to the one- and two-point correlation function calculations that we are interested in. Lastly, we use the minimal subtraction (MS) scheme for renormalization and leave our results in terms of an arbitrary renormalization parameter $\mu$.

%%%%%%%%%%%%%%%%%%%%%%%%%%%%%%%%%%%%%%%%%%%%%%%%%%

%------------------------------------------------
\section{The in-in formalism and initial state counterterms}
\label{sec:inin}
%------------------------------------------------

Finite-time correlation functions can be obtained by taking functional derivatives of the in-in generating functional,
\bea
    Z[J^+,J^-] & = & {\rm Tr} \[ \hat{\rho}(t_f) \]_{J^+,J^-} \nn \\
    & = & \int {\cal D}\phi^+(\vx,t) {\cal D}\phi^-(\vx,t) \bra{\phi^+} \hat{\rho}(t_0) \ket{\phi^-} {\rm exp} \bigg[ i\int \d^d x \big\{ {\cal L}[\phi^+] - {\cal L}[\phi^-] \nn \\
    & & \qquad + \ J^+\phi^+ - J^-\phi^- \big\} \bigg] \delta \[ \phi^+(\vx,t_f) - \phi^-(\vx,t_f) \] ,
\eea
with respect to the two sources $J^{\pm}(\vx,t)$ that are set to zero at the end of the calculation; see, for example, \cite{Berges:2004yj,Kamenev:2011}. The final time $t_f$ is chosen to be later than any other times of interest, and is also the turn-around point of the in-in contour; time integrals in the action thus run from $t_0$ to $t_f$. Plus and minus fields and sources lie on the forward and backward branches of the contour, respectively, and the $\delta$-function at the end imposes the boundary condition at the turn-around point. Note that $Z[J,J] = 1$ as long as the initial density operator is normalized.

We choose $\hat{\rho}(t_0)$ to be $|0\rangle \langle 0|$, $|0\rangle$ being the vacuum of the free theory, and write the initial density matrix as $\bra{\phi^+} \hat{\rho}(t_0) \ket{\phi^-} = N {\rm exp} \left\{ i{\cal S}[\phi^+, \phi^-; t_0] \right\}$, $N$ being a normalization constant chosen so that ${\rm Tr} \[ \hat{\rho}(t_0) \] = 1$. We can now proceed by making use of perturbation theory, writing the full action as $S[\phi] = S_0[\phi] + S_i[\phi]$, $S_0$ being the free part and $S_i$ the interaction part. On going to Fourier space,\footnote{We use the Fourier convention $f(\vx,t) \, = \, \int_{\vk} e^{i\vk \cdot \vx} f \big( \vk, t \big)$ with the shorthand $\int_{\vk} \equiv \int \frac{\d^{d-1} k}{(2\pi)^{d-1}}$ throughout this paper. The $\delta$-function is, therefore, given by $\int \d^{d-1} x \, e^{-i\vk \cdot \vx} = (2\pi)^{d-1} \delta^{d-1} \big( \vk \big)$. We further relegate Fourier indices to subscripts from now on for the ease of notation.} the generating functional can be written as
\bea
	Z[J^+,J^-] & = & {\rm exp} \left\{ iS_i \[-i\frac{\delta}{\delta J^+}\] - iS_i \[i\frac{\delta}{\delta J^-}\] + i\delta {\cal S} \[-i\frac{\delta}{\delta J^+}, i\frac{\delta}{\delta J^-}; t_0 \] \right\} \nn \\
	& & \quad \times \ {\rm exp} \[ - \frac{1}{2} \int_{t,t'} \int_{\vk} J^T_{\vk}(t) G_k(t,t') J_{-\vk}(t') \] ,
\label{eq:genfunc}
\eea
where we have used the shorthand $\int_t \equiv \int_{t_0}^{t_f} \d t$ for each time integral and defined $k = |\vk|$. We have also combined the two sources into a column vector,
\bea
	J_{\vk}(t) & = & \(
	\begin{array}{c}
		J_{\vk}^+(t) \\
		-J_{\vk}^-(t)
	\end{array} \) ,
\eea
with $J^T_{\vk}(t)$ being its transpose, and defined a $2 \times 2$ matrix of `Green's' functions,
\bea
	G_k(t,t') & = & \(
	\begin{array}{cc}
		G_k^{++}(t,t') & G_k^{+-}(t,t') \\
		G_k^{-+}(t,t') & G_k^{--}(t,t')
	\end{array} \) .
\label{eq:Gk2by2}
\eea
$G_k^{++}(t,t')$ and $G_k^{--}(t,t')$ here are genuinely Green's functions of the free theory whereas $G_k^{+-}(t,t')$ and $G_k^{-+}(t,t')$ are solutions of the homogeneous equations. We loosely refer to all four functions as Green's functions, however. The functional derivatives in $S_i$ and $\delta {\cal S}$, which we return to below, in eq.\ (\ref{eq:genfunc}), generate loop corrections similar to those in in-out calculations, except that the loops include all four functions $G_k^{\pm\pm}(t,t')$ rather than just the Feynman Green's function. The correlation function $\big\la \hPhi_{\vk}(t) \hPhi_{\vk'}(t') \big\ra$ is now given by
\bea
	\big\la \hPhi_{\vk}(t) \hPhi_{\vk'}(t') \big\ra & = & \( i \frac{\delta}{\delta J^-_{\vk}(t)} \) \( -i \frac{\delta}{\delta J^+_{\vk'}(t')} \) Z[J^+,J^-] \bigg|_{J^+ = \, J^- = \, 0} \, ,
	\label{eq:defgmp}
\eea
where we have used angular brackets to denote the expectation value in $|0\rangle$. The time-ordered two-point correlation is similarly obtained by taking both derivatives with respect to the source $J^+$. Note that any vacuum diagrams (disconnected diagrams without external sources) automatically vanish in in-in calculations, unlike in-out calculations, since $Z[0,0] = 1$ \cite{Weinberg:2005vy}. We will further ensure that the one-point correlation vanishes in our calculations below so that eq.\ (\ref{eq:defgmp}) matches with the connected two-point correlation. In the free theory, the two-point correlations simply give the Green's functions: $\big\la T \hPhi_{\vk}(t) \hPhi_{\vk'}(t') \big\ra_c = (2\pi)^{d-1} \delta^{d-1} \big( \vk+\vk' \big) G_k^{++}(t,t')$, where $T$ denotes time-ordering, $\big\la \hPhi_{\vk}(t) \hPhi_{\vk'}(t') \big\ra_c = (2\pi)^{d-1} \delta^{d-1} \big( \vk+\vk' \big) G_k^{-+}(t,t')$, $G_k^{+-}(t,t') = G_k^{-+*}(t,t')$, and $G_k^{--}(t,t') = G_k^{++*}(t,t')$. We can thus refer to loop corrections to the two-point correlations in an interacting theory as corrections to $G_k^{\pm\pm}(t,t')$.

Let us next write explicit expressions for $G_k^{\pm\pm}(t,t')$ since we will need them for our calculations in the following sections. One way to obtain them is to solve the Green's function equation with appropriate boundary conditions, which is especially useful when considering general initial states or non-unitary dynamics \cite{Agarwal:20xx}. For the choice of initial state here, we can simply use the fact that $G_k^{\pm\pm}(t,t')$ are two-point field correlators, and obtain them via canonical quantization. The free theory that we consider in this paper is that of a Klein-Gordon field, and therefore the interaction picture field $\hPhi_{\vk,I}(t)$ can be written as,
\bea
	\hPhi_{\vk,I}(t) & = & f_k^>(t) \hat{a}_{\vk} + f_k^<(t) \hat{a}_{-\vk}^{\dagger} \, ,
\label{eq:cq}
\eea
where $\hat{a}_{\vk}$ and $\hat{a}_{\vk}^{\dagger}$ are Schr\"odinger picture ladder operators defined at the initial time $t_0$, $f_k^>(t) = \frac{1}{\sqrt{2\omega_k}} e^{-i \omega_k (t-t_0)}$, $f_k^<(t) = f_k^{>*}(t)$, and $\omega_k = (k^2 + m^2)^{1/2}$. We also choose the following normalization of states: $\hat{a}_{\vk}^{\dagger} | 0 \ra = | \vk \ra$, $\la \vk | \vk' \ra = (2\pi)^{d-1} \delta^{d-1} \big( \vk-\vk' \big)$, and $\big[\hat{a}_{\vk}, \hat{a}_{\vk'}^{\dagger}\big] = (2\pi)^{d-1} \delta^{d-1} \big( \vk-\vk' \big)$. For the initial state $\hat{\rho}(t_0) = |0\ra \la0|$, the Green's function $G_k^{++}(t,t')$ and the function $G_k^{-+}(t,t')$ are then given by
\bea
	G_k^{++}(t,t') & = & f_k^>(t) f_k^<(t') \theta(t-t') + f_k^<(t) f_k^>(t') \theta(t'-t) \, ,
\label{eq:gpp} \\
	G_k^{-+}(t,t') & = & f_k^>(t) f_k^<(t') \, ,
\label{eq:gmp}
\eea
and, as mentioned earlier, the other two functions are obtained by taking complex conjugates of the above expressions.

Lastly, let us return to the function $\delta {\cal S}$ introduced in eq.\ (\ref{eq:genfunc}). Recall that we wrote the initial density matrix as $\bra{\phi^+} \hat{\rho}(t_0) \ket{\phi^-} = N {\rm exp} \left\{ i{\cal S}[\phi^+, \phi^-; t_0] \right\}$ and absorbed it into the definition of the Green's functions. The initial state that we have chosen is not an eigenstate of the full interacting Hamiltonian and is rather the ground state of just the free theory. We should then expect the act of switching on the interaction in an arbitrarily short timescale at $t_0$ to impact the initial state itself. In other words, we should expect the sudden transition at $t_0$ to be accompanied by particle creation. It is, therefore, not correct to leave the initial state independent of the dynamics and describe it solely with the function ${\cal S}$. Just as we added counterterms in the dynamics in eqs.\ (\ref{eq:lag3}) and (\ref{eq:lag4}), we thus add perturbative corrections to the initial state \cite{Calzetta:2008,Baacke:1997zz,Baacke:1999ia,Collins:2005nu,Collins:2014qna}. As we will see in the following sections, a Gaussian correction of the form \cite{Berges:2004yj,Agarwal:2012mq}
\bea
    \delta {\cal S}[\phi^+, \phi^-; t_0] & = & \frac{1}{2} \int_{\vk} \[ \delta A_k \phi^+_{\vk}(t_0) \phi^+_{-\vk}(t_0) - \delta A^*_k \phi^-_{\vk}(t_0) \phi^-_{-\vk}(t_0) + 2i \delta B_k \phi^+_{\vk}(t_0) \phi^-_{-\vk}(t_0) \] \quad \ \ 
\label{eq:isact}
\eea
will suffice, where $\delta A_k$ and $\delta B_k$ are initial state counterterms that will be chosen to cancel any UV divergences that can not be cancelled by counterterms in the dynamics. Note that while $\delta A_k$ is complex, $\delta B_k$ must be real so that the density operator is Hermitian.

%%%%%%%%%%%%%%%%%%%%%%%%%%%%%%%%%%%%%%%%%%%%%%%%%%

%------------------------------------------------
\section{Calculating finite-time correlations using in-in}
\label{sec:inincalc}
%------------------------------------------------

We will now use the in-in formalism in the next two subsections to calculate loop corrections to the unequal-time two-point correlator for the two interacting theories of eqs.\ (\ref{eq:lag3}) and (\ref{eq:lag4}).

%------------------------------------------------
\subsection{\texorpdfstring{$\lambda\phi^3$}{lp3} in 4D and 6D}
%------------------------------------------------

\begin{figure*}[!t]
\begin{center}
	\includegraphics[scale=0.6]{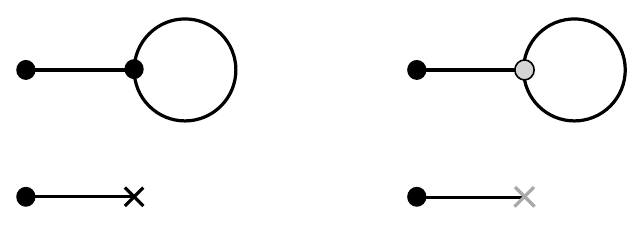}
	\caption{(Top) One loop contributions to the one-point function in a $\lambda\phi^3$ theory. (Bottom) The corresponding $Y$ counterterm diagrams.}
\label{fig:tp}
\end{center}
\end{figure*}

Let us first consider a $\lambda\phi^3$ interaction in $d$ dimensions, specializing to $d = 4$ or $d = 6$ later in the calculation. For a general $d$, we first calculate the one-loop correction to the one-point function, that contributes at ${\cal O}(\lambda)$, to fix the $Y$ counterterm and cancel any tadpole contributions going forward. The one-loop diagrams and counterterm diagrams are shown in fig.\ \ref{fig:tp}. The $2 \times 2$ matrix structure of Green's functions in eq.\ (\ref{eq:Gk2by2}) gives rise to the multiple similar-looking diagrams in the figure, where we use black dots and crosses to indicate vertices in the plus fields and grey dots and crosses for those in the minus fields, following the notation of \cite{Chen:2016nrs}. The total contribution from all diagrams is given by
\bea
    \big\la \hPhi_{\vk}(t) \big\ra & = & -i \frac{\delta}{\delta J^+_{\vk}(t)} Z[J^+,J^-] \bigg|_{J^+ = \, J^- = \, 0} \nn \\
    & = & -i(2\pi)^{d-1}\delta^{d-1} \big( \vk \big) \bigg[ \, \frac{\lambda}{2} \int_{t_1} \int_{\vp} \left\{ G_k^{++}(t,t_1) G_p^{++}(t_1,t_1) - G_k^{+-}(t,t_1) G_p^{--}(t_1,t_1) \right\} \nn \\
    & & \qquad + \ Y \int_{t_1} \left\{ G_k^{++}(t,t_1) - G_k^{+-}(t,t_1) \right\} \bigg] \, .
\label{eq:tp}
\eea
From eqs. (\ref{eq:gpp}) and (\ref{eq:gmp}), we see that the equal-time functions $G_k^{\pm\pm}(t,t)$ are all equal to $f_k^>(t) f_k^<(t)$ and, therefore, we can make $\big\la \hPhi_{\vk}(t) \big\ra$ vanish by choosing
\bea
    Y & = & \frac{\lambda\mu^{-\epsilon}}{2} \int_{\vp} f_p^>(t_1) f_p^<(t_1) \nn \\
    & = & \frac{\lambda\mu^{-\epsilon}}{2^d \pi^{(d-1)/2} \Gamma \[ (d-1)/2 \]} \int_0^{\infty} \d p \, \frac{p^{d-2}}{\sqrt{p^2+m^2}} \nn \\ \,
    & = & \frac{\lambda\mu^{-\epsilon} m^{d-2}}{2^{d+1} \pi^{d/2-1}} \frac{\csc(d\pi/2)}{\Gamma(d/2)}
\label{eq:linct}
\eea
for $d<2$. Above, we replaced $\lambda$ with $\lambda \mu^{4-d}$ or $\lambda \mu^{6-d}$ for $d = 4+\epsilon$ or $d = 6+\epsilon$, respectively, $\mu$ being a parameter with dimensions of mass, to keep the dimensions of $\lambda$ constant, and temporarily restored $m$. We can analytically continue the result to higher $d$ since we know that the interaction is renormalizable for both cases of interest. We see that $Y$ vanishes in the massless limit that we are interested in. It is also useful to note that $Y$ is independent of both $t$ and $\vk$ as befits Lorentz invariance and that adding this counterterm is equivalent to normal ordering the interaction, $\int \d^d x \, \frac{1}{3!} \lambda \hat{\Phi}^3 \rightarrow \int \d^d x \, \frac{1}{3!} \lambda {\,:} \hat{\Phi}^3 {\,:}$.

%========================= FIGURE 1 =========================
%============================================================

\begin{figure*}[!t]
\begin{center}
	\includegraphics[scale=0.59]{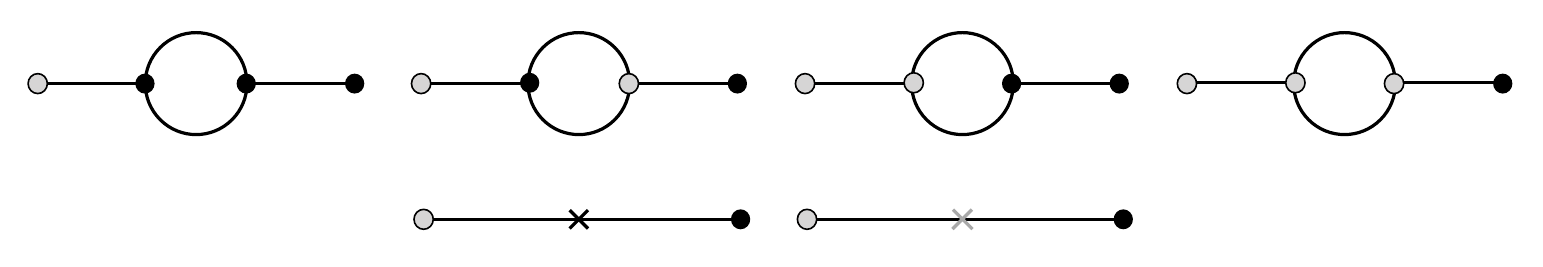}
	\caption{(Top) Connected one-loop contributions to the two-point correlation function in a $\lambda\phi^3$ theory. (Bottom) The corresponding $\delta_m$ and/or $\delta_r$ counterterm diagrams.}
\label{fig:2pphi3}
\end{center}
\end{figure*}

We next consider the one-loop correction to the two-point correlation, in particular the correction to $G_k^{-+}(t,t')$, that corresponds to the time-ordered correlator with $t > t'$. The connected one-loop diagrams that contribute at ${\cal O}(\lambda^2)$ and the $\delta_m$ and $\delta_r$ counterterm diagrams are shown in fig.\ \ref{fig:2pphi3}. Their contribution to $G_k^{-+}(t,t')$ is
\bea
    G_{k,{\rm 1-loop}}^{-+}(t,t') & = & G_{k, \lambda \phi^3}^{-+}(t,t') + G_{k, \delta_m \phi^2}^{-+}(t,t') + G_{k, \delta_r (\partial\phi)^2}^{-+}(t,t') \, ,
\label{Gkphi3full}
\eea
where
\bea
    G_{k, \lambda \phi^3}^{-+}(t,t') & = & -\frac{\lambda^2}{2} \int_{t_1, t_2} \int_{\vp} \Big[ G_k^{-+}(t,t_1) G_p^{++}(t_1,t_2) G_{|\vk-\vec{p}|}^{++}(t_1,t_2) G_k^{++}(t_2,t') \nn \\
    & & \qquad - \ G_k^{-+}(t,t_1) G_p^{+-}(t_1,t_2) G_{|\vk-\vec{p}|}^{+-}(t_1,t_2) G_k^{-+}(t_2,t') \nn \\
    & & \qquad - \ G_k^{--}(t,t_1) G_p^{-+}(t_1,t_2) G_{|\vk-\vec{p}|}^{-+}(t_1,t_2) G_k^{++}(t_2,t') \nn \\
    & & \qquad + \ G_k^{--}(t,t_1) G_p^{--}(t_1,t_2) G_{|\vk-\vec{p}|}^{--}(t_1,t_2) G_k^{-+}(t_2,t') \Big] \, ,
\label{eq:2plong}
\eea
and the two counterterm contributions are
\bea
    G_{k, \delta_m \phi^2}^{-+}(t,t') & = & -i\delta_m \int_{t_1} \[ G_k^{-+}(t,t_1) G_k^{++}(t_1,t') - G_k^{--}(t,t_1) G_k^{-+}(t_1,t') \] ,
\label{eq:2pdm} \\
    G_{k, \delta_r (\partial\phi)^2}^{-+}(t,t') & = & -i\delta_r k^2 \int_{t_1} \[ G_k^{-+}(t,t_1)G_k^{++}(t_1,t')-G_k^{--}(t,t_1)G_k^{-+}(t_1,t') \] \nn\\
    & & \qquad + \ i\delta_r \int_{t_1} \[ \dot{G}_k^{-+}(t,t_1) \dot{G}_k^{++}(t_1,t') -  \dot{G}_k^{--}(t,t_1) \dot{G}_k^{-+}(t_1,t') \] ,
\label{eq:2pdr}
\eea
with dots in the last line denoting partial derivatives with respect to $t_1$. The counterterm pieces are straightforward to calculate and so we write these first. On plugging in the expressions for the Green's functions from eqs.\ (\ref{eq:gpp}) and (\ref{eq:gmp}) and restricting to the massless case, we find that
\bea
    G_{k, \delta_m \phi^2}^{-+}(t,t') & = & \frac{\delta_m e^{-ik (t-t')}}{8k^3} \[ e^{2ik(t-t_0)} + e^{-2ik (t'-t_0)} - 2ik(t-t') - 2 \] ,
\label{eq:dmfin} \\
    G_{k, \delta_r (\partial\phi)^2}^{-+}(t,t') & = & \frac{\delta_r e^{-i k (t-t')}}{4k} \[ e^{2ik(t-t_0)} + e^{-2ik(t'-t_0)} - 2 \] .
\label{eq:drfin}
\eea
Note that these two contributions are the same for any interaction (and in any number of dimensions) and we, therefore, simply refer to them as needed to cancel the divergences in any of the interactions that we consider.

%========================= FIGURE 2 =========================
%============================================================

Let us now calculate the contribution from the cubic interaction in eq.\ (\ref{eq:2plong}). We first do a few manipulations to write this expression in a simpler form, that can then be calculated using {\it Mathematica}. We expand the Green's functions in the loop, $G_p$ and $G_{|\vk-\vp|}$, in terms of $\theta$-functions, for example, $G_p^{++}(t_1,t_2) = G_p^{-+}(t_1,t_2) \theta(t_1-t_2) + G_p^{+-}(t_1,t_2) \theta(t_2-t_1)$, and similarly for $G_p^{--}(t_1,t_2)$ and the $|\vk-\vp|$ pieces, and introduce $\theta$-functions in the second and third lines using $\theta(t_1-t_2) + \theta(t_2-t_1) = 1$. We next set the upper time integral $t_f$ to $t$, choosing $t > t'$. This allows us to rewrite eq. (\ref{eq:2plong}) as
\bea
    & & G_{k, \lambda \phi^3}^{-+}(t,t') \ = \ -\frac{\lambda^2}{2} \int_{t_1, t_2} \int_{\vp} \Big[ \nn \\
    & & \qquad \qquad G_p^{-+}(t_1,t_2) G_{|\vk-\vp|}^{-+}(t_1,t_2) G_k^{++}(t_2,t') \left\{ G_k^{-+}(t,t_1) - G_k^{+-}(t,t_1) \right\} \theta(t_1-t_2) \nn \\
    & & \qquad \qquad + \ G_p^{+-}(t_1,t_2) G_{|\vk-\vec{p}|}^{+-}(t_1,t_2) G_k^{-+}(t,t_1) \left\{ G_k^{++}(t_2,t') - G_k^{-+}(t_2,t') \right\} \theta(t_2-t_1) \nn \\
    & & \qquad \qquad + \ G_p^{-+}(t_1,t_2) G_{|\vk-\vec{p}|}^{-+}(t_1,t_2) G_k^{+-}(t,t_1) \left\{ G_k^{-+}(t_2,t')-G_k^{++}(t_2,t') \right\} \theta(t_2-t_1) \nn \\
    & & \qquad \qquad + \ G_p^{+-}(t_1,t_2) G_{|\vk-\vec{p}|}^{+-}(t_1,t_2) G_k^{-+}(t_2,t') \left\{ G_k^{+-}(t,t_1)-G_k^{-+}(t,t_1) \right\} \theta(t_1-t_2) \Big] \, , \qquad
\label{eq:2plongp}
\eea
where $\int_{t_i}$ now stands for $\int_{t_0}^t \d t_i$. Now since the integrals are symmetric in $t_1$ and $t_2$, we can make the substitution $t_1 \leftrightarrow t_2 $ in the second and third terms, so that all four terms are proportional to $\theta(t_1-t_2)$. We next expand the remaining $++$ Green's functions in terms of $\theta$-functions as well and insert $\theta(t_2-t') + \theta(t'-t_2) = 1$ in appropriate places to write eq.\ (\ref{eq:2plongp}) in the following form,
\bea
    G_{k, \lambda \phi^3}^{-+}(t,t') & = & G_{k, \lambda \phi^3, 1}^{-+}(t,t') + G_{k, \lambda \phi^3, 2}^{-+}(t,t') + G_{k, \lambda \phi^3, 2}^{-+}(t',t) \, ,
\label{eq:Gkphi34d}
\eea
where
\bea
    G_{k, \lambda \phi^3, 1}^{-+}(t,t') & = & -\frac{\lambda^2}{2} \int_{t'}^{t} \d t_1 \int_{t'}^{t_1} \d t_2 \int_{\vp} \Big[ G_p^{-+}(t_1,t_2) G_{|\vk-\vec{p}|}^{-+}(t_1,t_2) \big\{ G_k^{-+}(t,t_1) \nn \\
    & & \qquad - \ G_k^{+-}(t,t_1) \big\} \left\{ G_k^{-+}(t_2,t') - G_k^{+-}(t_2,t') \right\} \Big] \, ,
\label{eq:Ip1} \\
    G_{k, \lambda \phi^3, 2}^{-+}(t,t') & = & -\frac{\lambda^2}{2} \int_{t_0}^{t'} \d t_1 \int_{t_0}^{t_1} \d t_2 \int_{\vp} \Big[ G_p^{-+}(t_1,t_2) G_{|\vk-\vec{p}|}^{-+}(t_1,t_2) G_k^{-+}(t,t_2) \nn \\
    & & \qquad \times \left\{ G_k^{+-}(t_1,t') - G_k^{-+}(t_1,t') \right\} \Big] + \, {\rm c.c.} \, ,
\label{eq:Ip2}
\eea
with `c.c.' indicating the complex conjugate. We have essentially separated the contribution that is independent of the initial time $t_0$ from the one that depends on it. Before calculating the common loop integral, we will change $\int \d t_1 \int \d t_2$ to $\int \d \tau \int \d \Delta$, where for the time integrals in eq.\ (\ref{eq:Ip1}) we define $\tau = t_1 - t'$ and $\Delta = t_1 - t_2$ and for the time integrals in eq.\ (\ref{eq:Ip2}) we define $\tau = t_1 - t_0$ and $\Delta = t_1 - t_2$. We can also use the fact that the functions $G_k^{\pm\pm}(t,t')$ in Minkowski, and for our choice of initial state, only depend on the difference of times $t-t'$, so that we can write them as $G_k^{\pm\pm}(t-t')$. Then eqs.\ (\ref{eq:Ip1}) and (\ref{eq:Ip2}) become
\bea
    G_{k, \lambda \phi^3, 1}^{-+}(t,t') & = & -\frac{\lambda^2}{2} \int_{0}^{t-t'} \d \tau \int_{0}^{\tau} \d \Delta \int_{\vp} \Big[ G_p^{-+}(\Delta) G_{|\vk-\vec{p}|}^{-+}(\Delta) \big\{ G_k^{-+}(t-t'-\tau) \nn \\
    & & \qquad - \ G_k^{+-}(t-t'-\tau) \big\} \left\{ G_k^{-+}(\tau-\Delta) - G_k^{+-}(\tau-\Delta) \right\} \Big] \, ,
\label{eq:Ip12} \\
    G_{k, \lambda \phi^3, 2}^{-+}(t,t') & = & -\frac{\lambda^2}{2} \int_{0}^{t'-t_0} \d\tau \int_{0}^{\tau} \d \Delta \int_{\vp} \Big[ G_p^{-+}(\Delta) G_{|\vk-\vec{p}|}^{-+}(\Delta) G_k^{-+}(t-t_0-\tau+\Delta) \nn \\
    & & \qquad \times \left\{ G_k^{-+}(t'-t_0-\tau) - G_k^{+-}(t'-t_0-\tau) \right\} \Big] + \, {\rm c.c.} \, .
\label{eq:Ip22}
\eea
The lower limit of all time integrals is now zero, where we will find a pole later. We next calculate the loop integral first that is common to both contributions and given by
\bea
    & & I_{\lambda\phi^3}(k,\Delta) \ = \ \frac{\lambda^2}{2} \int_{\vp} G_p^{-+}(\Delta) G_{|\vk-\vec{p}|}^{-+}(\Delta) \nn \\
    & & \quad = \ \frac{\lambda^2}{2^{d+1} \pi^{d/2} \Gamma \( d/2-1 \)} \frac{1}{k} \int_0^{\infty} \d p \int_{|k-p|}^{k+p} \d q \, p^{d-4} \[ 1 - \frac{\(k^2 + p^2 - q^2 \)^2}{4 k^2 p^2} \]^{d/2-2} e^{-i(p+q)\Delta} \, , \ \ \nn \\
\label{eq:loopphi3}
\eea
where we have used $\int \d^d x = \frac{2\pi^{(d-1)/2}}{\Gamma [(d-1)/2]} \int_0^{\infty} \d x \int_{-1}^1 \d \cos\theta \, x^{d-1} \sin^{d-3}\theta$, defined $q = |\vec{k}-\vec{p}| = \( k^2 + p^2 - 2kp \cos\theta \)^{1/2}$, and changed the integral over $\cos\theta$ to $q$ using $\d \cos\theta = -\frac{q}{kp} \, \d q$. We have also restricted to the massless limit as mentioned earlier.

This is as far as we can get in the calculation of $G_{k,\lambda\phi^3}^{-+}$ (the one-loop correction to $G_k^{-+}$ from the $\lambda\phi^3$ interaction) without specifying the number of spacetime dimensions, and we will now specialize to $d = 4$ or $d = 6$.

%------------------------------------------------
\subsubsection*{(i) $\lambda\phi^3$ in 4D}
%------------------------------------------------

We first set $d = 4 + \epsilon$, where $\epsilon$ is infinitesimal and serves to regulate the integrals as before. In this case, the coupling $\lambda$ has the dimensions of mass. The loop integral in eq.\ (\ref{eq:loopphi3}) does not, in fact, need a regulator if we convert $\Delta$ into a Euclidean time coordinate, and simply leads to poles in the Euclidean $\Delta$. We thus define $\Delta = - i\bar{\Delta}$ and use $\epsilon$ to regulate the $\bar{\Delta}$ integral, setting $d = 4$ in the momentum integral. In other words, we take the fractional dimension to be solely in the time coordinate, leaving the number of spatial dimensions as an integer. Then eq.\ (\ref{eq:loopphi3}) becomes
\bea
    I_{\lambda\phi^3}(k,\bar{\Delta}) & = & \frac{\lambda^2}{32 \pi^2 k} \int_0^{\infty} \d p \int_{|k-p|}^{k+p} \d q \, e^{-(p+q)\bar{\Delta}} \nn \\
    & = & \frac{\lambda^2}{2} \frac{e^{-k\bar{\Delta}}}{16 \pi^2 \bar{\Delta}} \, .
\label{eq:loopdone}
\eea
We now use this expression in eqs.\ (\ref{eq:Ip12}) and (\ref{eq:Ip22}), converting the $\Delta$ to $\bar{\Delta}$ and leaving the $\tau$ as is, which then become
\bea
    G_{k, \lambda \phi^3, 1}^{-+}(t,t') & = & \frac{i\lambda^2\mu^{\epsilon}}{128 \pi^2 k^2} \frac{\pi^{(1+\epsilon)/2}}{\Gamma[(1+\epsilon)/2]} \int_{0}^{t-t'} \d\tau \int_{0}^{i\tau} \d\bar{\Delta} \, \bar{\Delta}^{\epsilon} \bigg[ \frac{ e^{-k\bar{\Delta}}}{\bar{\Delta}} \Big\{ e^{-ik(t-t'-\tau)} \nn \\
    & & \qquad - \ e^{ik(t-t'- \tau)} \Big\} \left\{ e^{-ik(\tau + i\bar{\Delta})} - e^{ik(\tau + i\bar{\Delta})} \right\} \bigg] \, , \\
    G_{k, \lambda \phi^3, 2}^{-+}(t,t') & = & \frac{i\lambda^2\mu^\epsilon}{128 \pi^2 k^2} \frac{\pi^{(1+\epsilon)/2}}{\Gamma[(1+\epsilon)/2]} \int_{0}^{t'-t_0} \d\tau \int_{0}^{i\tau} \d \bar{\Delta} \, \bar{\Delta}^{\epsilon} \bigg[ \frac{ e^{-k\bar{\Delta}}}{\bar{\Delta}} \, e^{-ik(t - t_0 - \tau - i\bar{\Delta})} \nn \\
    & & \qquad \times \left\{ e^{-ik(t'-t_0-\tau)} - e^{ik(t'-t_0-\tau)} \right\} \Big] + \, {\rm c.c.} \, ,
\eea
where, as before, $\mu$ is a parameter with dimensions of mass, introduced to keep the dimensions of $\lambda$ fixed. These integrals are straightforward to calculate. The result can be plugged into eq.\ (\ref{eq:Gkphi34d}) to obtain the contribution from the interaction to the one-loop two-point correlator in eq.\ (\ref{Gkphi3full}). We can then perform a series expansion around $\epsilon = 0$ to separate the finite piece from the UV-divergent piece that must be canceled by an appropriate choice of counterterms.

Let us first look at the finite (i.e. independent of $\epsilon$) part of $G_{k,\lambda\phi^3}^{-+}(t,t')$ since this is the main result of the current section. We will call this the one-loop correction to $G_k^{-+}(t,t')$ since all UV-divergent terms can be canceled by counterterms, as we will see later. We find that 
\bea
    & & G_{k,{\rm 1-loop}}^{-+}(t,t') \nn \\
    & & \ \ = \ -\frac{i\lambda^2e^{-i k (t-t')}(t-t')}{128 k^2 \pi^2} \Bigg[ 1 - \ln \( \frac{2\pi\mu^2(t-t')}{k} \) - {\rm Ei} \left\{ 2ik(t-t_0) \right\}+i\frac{\pi}{2}\nn \\
    & & \qquad \quad + \ e^{2ik(t-t')}\Big({\rm Ei} \left\{ -2ik (t-t_0) \right\}-{\rm Ei} \left\{ -2ik (t-t') \right\}\Big) \Bigg] \nn \\
    & & \qquad - \frac{i\lambda^2 e^{-i k (t-t')}(t'-t_0)}{128 k^2 \pi^2} \Bigg[ 2i\pi + {\rm Ei} \left\{ -2ik(t'-t_0) \right\}-{\rm Ei} \left\{ 2ik(t-t_0) \right\}\nn \\
    & & \qquad \quad + \ e^{2ik(t-t')} \Big({\rm Ei} \left\{ -2ik(t-t_0) \right\}-{\rm Ei} \left\{ 2ik(t'-t_0) \right\} +2i\pi\Big) \Bigg] \nn \\
    & & \qquad + \frac{\lambda^2e^{-i k (t-t')}}{256 k^3\pi^2} \Bigg[3 + {\rm Ei} \left\{ 2ik(t-t_0) \right\}+{\rm Ei} \left\{ -2ik(t'-t_0) \right\}  \nn \\
    & & \qquad \quad + \ 2 \ln \( \frac{4i\pi\mu^2(t-t')}{2k} \) + e^{2 ik(t-t')}\Big(1 -2{\rm Ei} \left\{ -2ik(t-t') \right\}-i\pi +{\rm Ei} \left\{ 2ik(t'-t_0) \right\}\nn \\
    & & \qquad \quad + \ {\rm Ei} \left\{ -2ik(t-t_0) \right\}\Big) -e^{i k (t-t')} \Big(2\cos \left\{ k (t+t'-2t_0) \right\} \left(2+\gamma \right) \nn \\
    & & \qquad \quad + \ 2\cos\left\{ k(t+t'-2t_0) \right\} \ln \left\{4\pi \mu^2(t-t_0)(t'-t_0) \right\} + 2\pi \sin \left\{ k (t+t'-2t_0) \right\}\Big)\Bigg] \, , \nn \\
\label{eq:2pfinite}
\eea
where ${\rm Ei}(z) = \int_{-z}^{\infty} \d t \, \frac{e^{-t}}{t}$ is the (principal value of the) exponential integral, $\gamma$ is the Euler-Mascheroni constant, and we have used the MS renormalization scheme (we return to this point below). There are several interesting things to note about this expression. First, it vanishes in the limit $t = t' = t_0$, as expected from our initial condition. Second, it is time translation invariant as long as we shift all times, including the initial time, by the same amount. Third, there is no IR issue despite the factor of $1/k^3$ outside, since the $k \rightarrow 0$ limit of eq.\ (\ref{eq:2pfinite}) in fact goes as $1/k$, similar to the free theory. Fourth, if we take the equal-time limit $t = t'$, further take the late-time limit so that $t-t_0$ is bigger than any timescale in the problem, $\lambda(t-t_0) \gg 1$, $k(t-t_0) \gg 1$, and $\mu(t-t_0) \gg 1$, and drop oscillatory terms by averaging over the late time interval, then we find that
\bea
    \lim_{t \gg t_0} G_{k,{\rm 1-loop}}^{-+}(t,t) & = & \frac{\lambda^2}{128k^3\pi^2} \[ 2 - \gamma - \ln\( \frac{\pi \mu^2}{k^2} \) \] ,
\eea
which does not exhibit any secular growth, consistent with the expectation that the (free theory) vacuum is stable under a $\lambda\phi^3$ interaction in Minkowski \cite{Chaykov:2022zro}. And fifth, the late-time limit, such that all time differences $T = t-t_0$, $t'-t_0$, and $t-t'$ satisfy $\lambda T \gg 1$, $kT \gg 1$, and $\mu T \gg 1$, {\it does} exhibit secular growth with
\bea
    \lim_{t \gg t' \gg t_0} G_{k,{\rm 1-loop}}^{-+}(t,t') & = & \frac{\lambda^2 (t-t') e^{-ik(t-t')}}{256k^2\pi^2} \[ - 2i - \pi + 2i \ln\( \frac{2\pi\mu^2(t-t')}{k} \) \] .
\label{eq:phi3lt}
\eea
This is consistent with what we find in our companion paper \cite{Chaykov:2022zro}, where we use an in-out approach to directly obtain the late-time result, with differences that can be absorbed into the counterterms. As shown there, the Weisskopf-Wigner (WW) resummation method resums the late-time secular growth into exponential decay, consistent with the expectation that while the vacuum is stable in Minkowski, the one-particle state decays. Also note that the expression in eq.\ (\ref{eq:phi3lt}) goes as $
(1/k^2)\ln k$ in the $k \rightarrow 0$ limit, which still does not lead to an IR issue on converting to real space. 

Let us next consider the UV-divergent piece of $G_{k,\lambda\phi^3}^{-+}(t,t')$,
\bea
    G_{k,\lambda\phi^3,\epsilon}^{-+}(t,t') & = & \frac{\lambda^2 e^{-ik(t-t')}}{128\pi^2 k^3\epsilon} \[2 + 2ik(t-t')-  e^{2ik(t-t_0)} -e^{-2ik(t'-t_0)} \] .
\eea
Comparing this to the $\delta_m$ and $\delta_r$ contributions in eqs.\ (\ref{eq:dmfin}) and (\ref{eq:drfin}), we see that the UV-divergence can be fully canceled by choosing the counterterms to be
\bea
    \delta_m \ = \ \frac{\lambda^2}{16\pi^2\epsilon} \, , \quad \delta_r \ = \ 0 \, ,
\label{eq:ctphi34d}
\eea
where we have used the MS renormalization scheme. Note that we do not need to add counterterms in the initial state, and so we can take $\delta A_k$ and $\delta B_k$ introduced in section\ \ref{sec:inin} to both be zero.

%------------------------------------------------
\subsubsection*{(ii) $\lambda\phi^3$ in 6D}
%------------------------------------------------

We next set $d = 6 + \epsilon$, so that $\lambda$ is dimensionless. As in the 4D case, we define $\Delta = -i\bar{\Delta}$ and use $\epsilon$ to regulate the $\bar{\Delta}$ integral, setting $d = 6$ in the momentum integral. Now eq.\ (\ref{eq:loopphi3}) becomes
\bea
    I_{\lambda\phi^3}(k,\bar{\Delta}) & = & \frac{\lambda^2}{128 \pi^3 k} \int_0^{\infty} \d p \int_{|k-p|}^{k+p} \d q \[ p^2 - \frac{\( k^2 + p^2 - q^2 \)^2}{4 k^2} \] e^{-(p+q)\bar{\Delta}} \nn \\
    & = & \frac{\lambda^2}{2} \frac{e^{-k \bar{\Delta}} \( 1 + k\bar{\Delta} \)}{192 \pi^3 \bar{\Delta}^3} \, .
\label{eq:loopdone6}
\eea
Using this expression in eqs.\ (\ref{eq:Ip12}) and (\ref{eq:Ip22}), plugging the result into eq.\ (\ref{eq:Gkphi34d}), and series expanding around $\epsilon= 0$ (the integrals converge for $\epsilon > 2$ but we can analytically continue the result to smaller $\epsilon$), once again gives us the finite and UV-divergent pieces of the one-loop two-point correlator in eq.\ (\ref{Gkphi3full}). The finite part of the one-loop correction is now given by
\bea
    & & G_{k,{\rm 1-loop}}^{-+}(t,t') \ = \ \frac{\lambda^2 e^{-i k (t-t')}}{3072 \pi^3 k} \Bigg[ \gamma - i\pi  -2\ln \left(\frac{2\pi^{3/2}\mu^3(t-t')}{k^2 }\right)-2\text{Ei} \left\{ -2 i k (t-t_0) \right\}-2 \nn \\
    & & \qquad - \ 2\text{Ei} \left\{ -2 i k (t'-t_0) \right\} + e^{2 i k (t-t')}\Bigg( \gamma + 2i\pi+2\text{Ei} \left\{ -2 i k (t-t') \right\} -2\text{Ei} \left\{- 2 i k (t-t_0) \right\} \nn \\
    & & \qquad - \ 2\text{Ei} \left\{ 2 i k (t'-t_0) \right\} -\ln \left(\frac{\pi\mu^2}{k^2 }\right)\Bigg) \Bigg] \, ,
\label{eq:2pfinite6}
\eea
where $\mu$ is again a parameter with dimensions of mass, introduced to keep the dimensions of $\lambda$ fixed. This expression shares the first four features mentioned after eq.\ (\ref{eq:2pfinite}), except for a subtlety when taking the $t=t'=t_0$ limit. Once we set $t=t'$ and take the limit as $t \rightarrow t_0$, we find a $\ln \{ \mu(t-t_0) \}$ divergence. This can be removed by choosing $\mu = 1/(t-t_0)$, this being the only scale left in the problem, and is likely related to the simultaneous appearance of non-zero initial state counterterms that we discuss below. Also, to show the equal-time late-time limit explicitly, if we set $t = t'$, take $k(t-t_0) \gg 1$ and $\mu(t-t_0) \gg 1$ we find that
\bea
    \lim_{t \gg t_0} G_{k,{\rm 1-loop}}^{-+}(t,t) & = & \frac{\lambda^2}{768k\pi^3} \[ \gamma - \ln\( \frac{\pi \mu^2}{k^2} \) \] ,
\eea
which again does not exhibit any secular growth. The late-time limit of eq.\ (\ref{eq:2pfinite6}), such that all time differences $T = t-t_0$, $t'-t_0$, and $t-t'$ satisfy $kT \gg 1$ and $\mu T \gg 1$, on the other hand, now exhibits secular growth that is logarithmic rather than linear in $t-t'$,
\bea
    \lim_{t \gg t' \gg t_0} G_{k,{\rm 1-loop}}^{-+}(t,t') & = & -\frac{\lambda^2e^{-ik(t-t')}}{1536\pi^3k}\ln\left\{ \mu(t-t') \right\} .
\eea
This is also consistent with what we find in our companion paper \cite{Chaykov:2022zro} and, as shown there, the result can be resummed to yield a polynomial decay.

Lastly, let us look at the UV-divergent piece of $G_{k,\lambda\phi^3}^{-+}(t,t')$,
\bea
    G_{k,\lambda\phi^3,\epsilon}^{-+}(t,t') & = & -\frac{\lambda^2 e^{-ik(t-t')}}{1536 \pi^3 k\epsilon} \[ e^{2ik(t-t')} + 3 \] .
\label{eq:Gkphi3epsilon}
\eea
Comparing this to the $\delta_m$ and $\delta_r$ contributions in eqs.\ (\ref{eq:dmfin}) and (\ref{eq:drfin}), we see that the UV-divergence is not fully canceled by these counterterms. Therefore, we need additional counterterms in the initial state in this case. To calculate the contributions from $\delta A_k$ and $\delta B_k$, we can treat the initial state action in eq.\ (\ref{eq:isact}) as any other counterterm action, with the only difference that it is defined at the initial time. Then following steps very similar to those outlined in eqs.\ (\ref{eq:2pdm})-(\ref{eq:drfin}), we find that the additional counterterm contributions to $G_{k,{\rm 1-loop}}^{-+}(t,t')$ are
\bea
    G_{k, \delta A_k \phi^2}^{-+}(t,t') & = & \frac{i\delta A_k^* e^{ik(t+t'-2t_0)}}{4k^2} - \frac{i\delta A_k e^{-ik(t+t'-2t_0)}}{4k^2} \, ,
\label{eq:isctA} \\
    G_{k, \delta B_k \phi^2}^{-+}(t,t') & = & -\frac{\delta B_k \cos\[ k(t-t') \]}{2k^2} \, .
\label{eq:isctB}
\eea
Now we can start with fixing $\delta B_k$ so that it cancels the $e^{ik(t-t')}$ part of the UV divergence in eq.\ (\ref{eq:Gkphi3epsilon}), since this is the only counterterm that contains such a term. We can next choose $\delta_r$ so that it cancels the remaining piece proportional to $e^{-ik(t-t')}$ and finally choose $\delta A_k$ to cancel any remaining divergences. This leads to the following choice of counterterms,
\bea
    \delta_m \ = \ 0 \, , \quad \delta_r \ = \ -\frac{\lambda^2}{384 \pi^3 \epsilon} \, , \quad \delta A_k \ = \ \frac{i\lambda^2k}{384 \pi^3 \epsilon} \, , \quad \delta B_k \ = \ -\frac{\lambda^2 k}{384 \pi^3 \epsilon} \, .
\label{eq:ctphi36d}
\eea
An interesting point to note is that $\delta B_k$ is non-zero. Looking back at eq.\ (\ref{eq:isact}), we see that this couples the $\phi^+$ and $\phi^-$ fields at the initial time, and therefore leads to mixing in the initial state. This is also confirmed by calculating the purity of the state, that is indeed less than unity if $\delta B_k$ is non-zero (the zeroth order $B_k$ vanishes for our choice of initial state).

%------------------------------------------------
\subsection{\texorpdfstring{$\lambda\phi^4$}{lp4} in 4D}
%------------------------------------------------

\begin{figure*}[!t]
\begin{center}
	\includegraphics[scale=0.6]{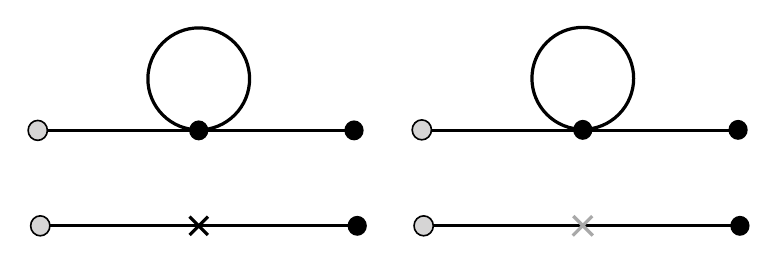}
	\caption{(Top) One-loop contributions to the two-point correlation function in a $\lambda\phi^4$ theory. (Bottom) The corresponding $\delta_m$ and/or $\delta_r$ counterterm diagrams.}
\label{fig:2pphi4}
\end{center}
\end{figure*}

Let us next consider a $\lambda\phi^4$ interaction in $d$ dimensions, specializing to $d = 4$ later in the calculation. Since the $\lambda\phi^4$ interaction does not generate a one-point expectation value, we can directly look at the two-point correlation, specifically $G_k^{-+}(t,t')$ with $t > t'$. We first calculate the one-loop correction to it that contributes at ${\cal O}(\lambda)$. The one-loop diagrams and the $\delta_m$ and $\delta_r$ counterterm diagrams are shown in fig.\ \ref{fig:2pphi4} and their contribution to $G_k^{-+}(t,t')$ is
\bea
    G_{k,{\rm 1-loop}}^{-+}(t,t') & = & G_{k, \lambda \phi^4}^{-+}(t,t') + G_{k, \delta_m \phi^2}^{-+}(t,t') + G_{k, \delta_r (\partial\phi)^2}^{-+}(t,t') \, ,
\label{Gkphi41loopfull}
\eea
where
\bea
    G_{k, \lambda \phi^4}^{-+}(t,t') & = & \frac{i\lambda}{2} \int_{t_1} \int_{\vp} \Big[ G_k^{-+}(t,t_1) G_p^{++}(t_1,t_1) G_k^{++}(t_1,t') \nn \\
    & & \qquad - \ G_k^{--}(t,t_1) G_p^{--}(t_1,t_1) G_k^{-+}(t_1,t') \Big] \, ,
\label{eq:phi41loop}
\eea
and the two counterterm contributions are the same as those of the previous subsection. On using the fact that all $G_k^{\pm\pm}(t,t)$ are the same, we see that the loop integral here is identical to that in eq.\ (\ref{eq:linct}) for the $Y$ counterterm in the cubic theory. As found there, the loop integral vanishes in the massless limit that we are working in, and so eq.\ (\ref{eq:phi41loop}) also vanishes in this limit.

%========================= FIGURE 3 =========================
%============================================================

\begin{figure*}[!t]
\begin{center}
	\includegraphics[scale=0.58]{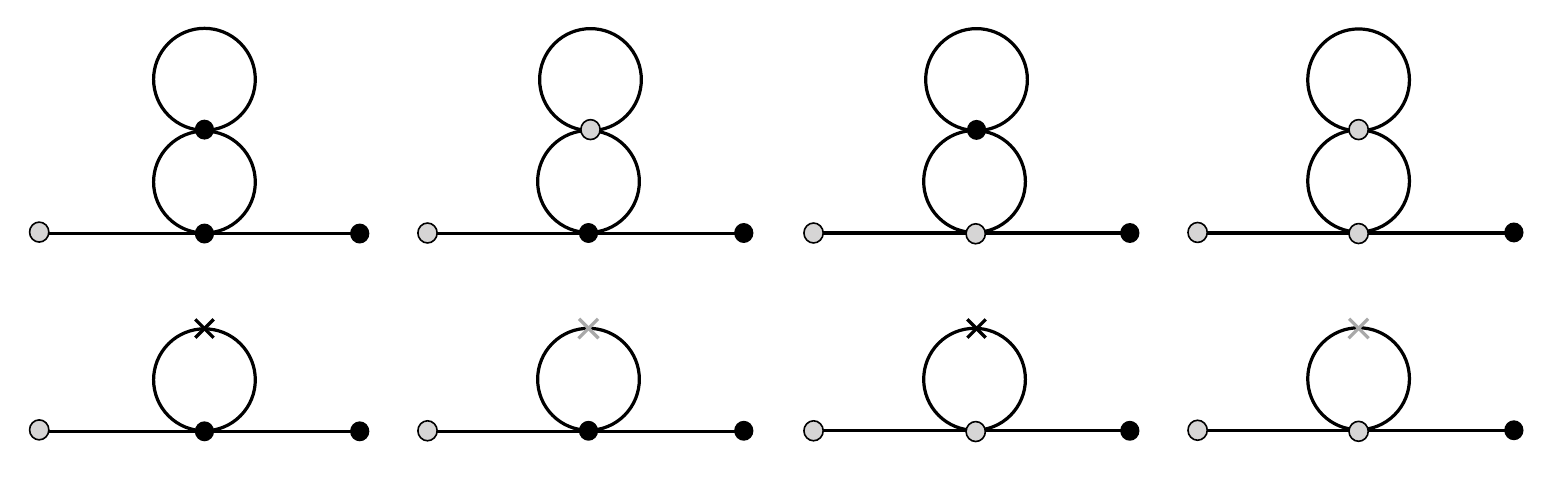}
	\caption{(Top) Connected two-loop contributions of the `snowman' type to the two-point correlation function in a $\lambda\phi^4$ theory. (Bottom) The corresponding $\delta_m$ and/or $\delta_r$ counterterm diagrams.}
\label{fig:2psm}
\end{center}
\end{figure*}

\begin{figure*}[!t]
\begin{center}
	\includegraphics[scale=0.59]{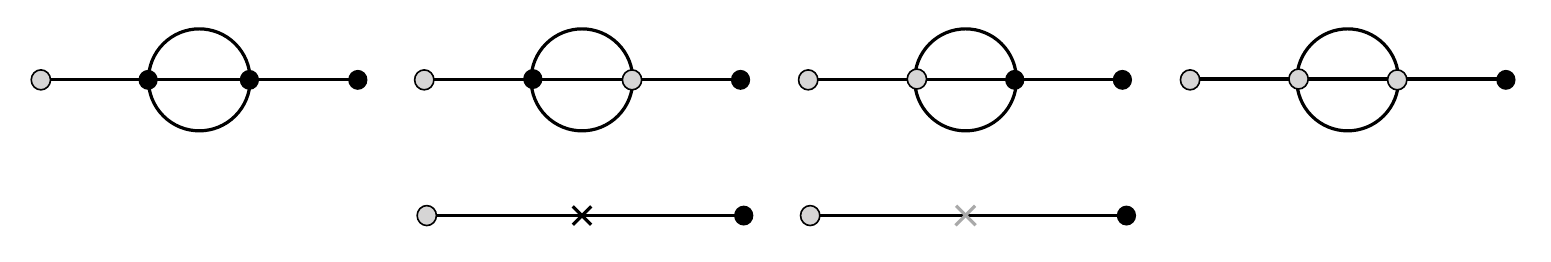}
	\caption{(Top) Connected two-loop contributions of the `sunset' type to the two-point correlation function in a $\lambda\phi^4$ theory. (Bottom) The corresponding $\delta_m$ and/or $\delta_r$ counterterm diagrams.}
\label{fig:2pss}
\end{center}
\end{figure*}

Since the one-loop correction vanishes, we also calculate the two-loop correction to the two-point correlation. There are two types of two-loop diagrams, both contributing at ${\cal O}(\lambda^2)$. The first set of diagrams, that we refer to as the snowman (`sm') diagrams, along with the corresponding $\delta_m$ and $\delta_r$ counterterm diagrams are shown in fig. \ref{fig:2psm} and the second set of diagrams, that we refer to as the sunset (`ss') diagrams, along with the corresponding $\delta_m$ and $\delta_r$ counterterm diagrams are shown in fig. \ref{fig:2pss}. Their full contribution to $G_k^{-+}(t,t')$ is
\bea
    G_{k,{\rm 2-loop}}^{-+}(t,t') & = & G_{k, \lambda\phi^4, {\rm sm}}^{-+}(t,t') + G_{k, \delta_m \phi^2,{\rm sm}}^{-+}(t,t') + G_{k, \delta_r (\partial\phi)^2,{\rm sm}}^{-+}(t,t') \nn \\
    & & \quad + \ G_{k, \lambda\phi^4, {\rm ss}}^{-+}(t,t') + G_{k, \delta_m \phi^2,{\rm ss}}^{-+}(t,t') + G_{k, \delta_r (\partial\phi)^2,{\rm ss}}^{-+}(t,t') \, ,
\label{Gkphi42loopfull}
\eea
where the first line corresponds to the diagrams in fig.\ \ref{fig:2psm} and the second to those in fig.\ \ref{fig:2pss}. The contribution from the snowman diagrams is
\bea
    G_{k, \lambda\phi^4, {\rm sm}}^{-+}(t,t') & = & -\frac{\lambda^2}{2} \int_{t_1, t_2} \int_{\vp_1, \vp_2} \Big[ G_k^{-+}(t,t_1) G_{p_1}^{++}(t_1,t_2) G_{p_1}^{++}(t_1,t_2) G_{p_2}^{++}(t_2,t_2) G_k^{++}(t_1,t') \nn \\
    & & \qquad - \ G_k^{-+}(t,t_1) G_{p_1}^{+-}(t_1,t_2) G_{p_1}^{+-}(t_1,t_2) G_{p_2}^{--}(t_2,t_2) G_k^{++}(t_1,t') \nn \\
    & & \qquad - \ G_k^{--}(t,t_1) G_{p_1}^{-+}(t_1,t_2) G_{p_1}^{-+}(t_1,t_2) G_{p_2}^{++}(t_2,t_2) G_k^{-+}(t_1,t') \nn \\
    & & \qquad + \ G_k^{--}(t,t_1) G_{p_1}^{--}(t_1,t_2) G_{p_1}^{--}(t_1,t_2) G_{p_2}^{--}(t_2,t_2) G_k^{-+}(t_1,t') \Big] \, .
\label{eq:p42psm}
\eea
The $\vp_2$ integrals here are of the same form as encountered in the one-loop correction and eq.\ (\ref{eq:linct}), and therefore, the snowman contribution also vanishes in the massless limit.

%========================= FIGURE 4 =========================
%============================================================

The only nonzero correction to the two-point correlation comes from the sunset diagrams and is given by
\bea
    & & G_{k, \lambda\phi^4, {\rm ss}}^{-+}(t,t') \ = \ -\frac{\lambda^2}{6} \int_{t_1, t_2} \int_{\vp_1, \vp_2} \Big[ \nn \\
    & & \qquad \qquad G_k^{-+}(t,t_1) G_{p_1}^{++}(t_1,t_2 )G_{p_2}^{++}(t_1,t_2) G_{|\vec{k}-\vec{p}_1-\vec{p_2}|}^{++}(t_1,t_2) G_k^{++}(t_2,t') \nn \\
    & & \qquad \qquad - \ G_k^{-+}(t,t_1) G_{p_1}^{+-}(t_1,t_2) G_{p_2}^{+-}(t_1,t_2) G_{|\vec{k}-\vec{p}_1-\vec{p_2}|}^{+-}(t_1,t_2) G_k^{-+}(t_2,t') \nn \\
    & & \qquad \qquad - \ G_k^{--}(t,t_1) G_{p_1}^{-+}(t_1,t_2) G_{p_2}^{-+}(t_1,t_2) G_{|\vec{k}-\vec{p}_1-\vec{p_2}|}^{-+}(t_1,t_2) G_k^{++}(t_2,t') \nn \\
    & & \qquad \qquad + \ G_k^{--}(t,t_1) G_{p_1}^{--}(t_1,t_2) G_{p_2}^{--}(t_1,t_2) G_{|\vec{k}-\vec{p}_1-\vec{p_2}|}^{--}(t_1,t_2) G_k^{-+}(t_2,t') \Big] \, .
\label{eq:p42pss}
\eea
To solve this, we first perform the same manipulations as those in the previous subsection. In fact, we can borrow many of the expressions that we had there, noticing that eq.\ (\ref{eq:p42pss}) can simply be obtained from eq.\ (\ref{eq:2plong}) by making the replacement
\bea
    -\frac{\lambda^2}{2} \int_{\vp} G_p^{ab}(t_1,t_2) G_{|\vec{k}-\vec{p}|}^{ab}(t_1,t_2) \rightarrow -\frac{\lambda^2}{6} \int_{\vp_1} \int_{\vp_2} G_{p_1}^{ab}(t_1,t_2) G_{p_2}^{ab}(t_1,t_2) G_{|\vec{k}-\vec{p}_1-\vec{p}_2|}^{ab}(t_1,t_2) \, . \quad \ \ 
\label{eq:p3top4}
\eea
for any $\pm$ indices $a$ and $b$. We can, therefore, write the sunset contribution in a form analogous to eq.\ (\ref{eq:Gkphi34d}) of the previous subsection,
\bea
    G_{k, \lambda\phi^4, {\rm ss}}^{-+}(t,t') & = & G_{k, \lambda\phi^4, {\rm ss}, 1}^{-+}(t,t') + G_{k, \lambda\phi^4, {\rm ss}, 2}^{-+}(t,t') + G_{k, \lambda\phi^4, {\rm ss}, 2}^{-+}(t',t) \, ,
\label{eq:Gkphi4}
\eea
where $G_{k, \lambda\phi^4, {\rm ss}, 1}^{-+}(t,t')$ and $G_{k, \lambda\phi^4, {\rm ss}, 2}^{-+}(t,t')$, given below for clarity, are obtained from eqs.\ (\ref{eq:Ip1}) and (\ref{eq:Ip2}), respectively, on making the replacement of eq.\ (\ref{eq:p3top4}),
\bea
    G_{k, \lambda\phi^4, {\rm ss}, 1}^{-+}(t,t') & = & -\frac{\lambda^2}{6} \int_{t'}^{t} \d t_1 \int_{t'}^{t_1} \d t_2 \int_{\vp_1} \int_{\vp_2} \Big[ G_{p_1}^{-+}(t_1,t_2) G_{p_2}^{-+}(t_1,t_2) G_{|\vec{k}-\vec{p}_1-\vec{p}_2|}^{-+}(t_1,t_2) \nn \\
    & & \qquad \times \ \big\{ G_k^{-+}(t,t_1) - G_k^{+-}(t,t_1) \big\} \left\{ G_k^{-+}(t_2,t') - G_k^{+-}(t_2,t') \right\} \Big] \, , \\
    G_{k, \lambda\phi^4, {\rm ss}, 2}^{-+}(t,t') & = & -\frac{\lambda^2}{6} \int_{t_0}^{t'} \d t_1 \int_{t_0}^{t_1} \d t_2 \int_{\vp_1} \int_{\vp_2} \Big[ G_{p_1}^{-+}(t_1,t_2) G_{p_2}^{-+}(t_1,t_2) G_{|\vec{k}-\vec{p}_1-\vec{p}_2|}^{-+}(t_1,t_2) \nn \\
    & & \qquad \times \ G_k^{-+}(t,t_2) \left\{ G_k^{+-}(t_1,t') - G_k^{-+}(t_1,t') \right\} \Big] + \, {\rm c.c.} \, .
\eea
We can change $\int \d t_1 \int \d t_2$ to $\int \d \tau \int \d \Delta$ and write the Green's functions in terms of a single time variable as before. Also setting the mass to zero and the number of spatial dimensions to three, the loop integral that needs to be solved is then given by
\bea
    I_{\lambda\phi^4}(k,\Delta) & = & \frac{\lambda^2}{6} \int_{\vp_1} \int_{\vp_2} G_{p_1}^{-+}(\Delta) G_{p_2}^{-+}(\Delta) G_{|\vec{k}-\vec{p}_1-\vec{p}_2|}^{-+}(\Delta) \nn \\
    & = & \frac{\lambda^2}{768\pi^4 k} \int_{0}^{\infty} \d p_1 \int_{|k-p_1|}^{k+p_1} \d q_1 \int_{0}^{\infty} \d p_2 \int_{|q_1-p_2|}^{q_1+p_2} \d q_2 \, e^{-i \(p_1 + p_2 + q_2 \) \Delta} \, ,
\label{eq:loopphi4}
\eea
where we have defined $\vec{q}_1 = \vk - \vp_1$ and $\vec{q}_2 = \vec{q}_1 - \vp_2$, so that $q_1 = \( k^2 + p_1^2 - 2kp_1 \cos\theta_1 \)^{1/2}$ and $q_2 = \( q_1^2 + p_2^2 - 2q_1 p_2 \cos\theta_2 \)^{1/2}$, and changed the integrals over $\cos\theta_1$ and $\cos\theta_2$ to $q_1$ and $q_2$ using $\d \cos\theta_1 = -\frac{q_1}{kp_1} \, \d q_1$ and $\d \cos\theta_2 = -\frac{q_2}{q_1 p_2} \, \d q_2$, respectively.

%========================= FIGURE 5 =========================
%============================================================

In order to regulate the integrals, we define a Euclidean time coordinate $\Delta = -i\bar{\Delta}$ and set $d = 4 + \epsilon$, absorbing the fractional dimension fully into the time coordinate, so that the number of spatial dimensions is still three. The integrals in eq.\ (\ref{eq:loopphi4}) can then be performed easily and we find that
\bea
    I_{\lambda\phi^4}(k,\bar{\Delta}) & = & \frac{\lambda^2}{6} \frac{e^{-k\bar{\Delta}} \( 1 + k\bar{\Delta} \)}{256 \pi^4 \bar{\Delta}^3} \, .
\label{eq:p4loopdone}
\eea
Notice that this is exactly the same (up to a constant) as what we found for $\lambda\phi^3$ in 6D, in eq.\ (\ref{eq:loopdone6}), with both interactions sharing the fact that $\lambda$ is dimensionless. We can, therefore, directly write the contribution from the sunset diagrams using eq.\ (\ref{eq:2pfinite6}) and we find that the finite part of the two-loop correction to the two-point correlation is given by
\bea
    & & G_{k,{\rm 2-loop}}^{-+}(t,t') \ = \ \frac{\lambda^2 e^{-i k (t-t')}}{12288 \pi^4 k} \Bigg[ \gamma - i\pi  -2\ln \left(\frac{2\pi^{3/2}\mu^3(t-t')}{k^2 }\right)-2\text{Ei} \left\{ -2 i k (t-t_0) \right\}-2 \nn \\
    & & \qquad - \ 2\text{Ei} \left\{ -2 i k (t'-t_0) \right\} + e^{2 i k (t-t')}\Bigg( \gamma + 2i\pi+2\text{Ei} \left\{ -2 i k (t-t') \right\} -2\text{Ei} \left\{- 2 i k (t-t_0) \right\} \nn \\
    & & \qquad - \ 2\text{Ei} \left\{ 2 i k (t'-t_0) \right\} -\ln \left(\frac{\pi\mu^2}{k^2 }\right)\Bigg) \Bigg] \, .
\label{eq:2pfinite4}
\eea
The qualitative behavior of the loop correction matches what we found for $\lambda\phi^3$ in 6D, eq.\ (\ref{eq:2pfinite6}). Similar to what we found there, the equal-time late-time limit with $t = t'$, $k(t-t_0) \gg 1$, and $\mu(t-t_0) \gg 1$, does not exhibit any secular growth,
\bea
    \lim_{t \gg t_0} G_{k,{\rm 1-loop}}^{-+}(t,t) & = & \frac{\lambda^2}{3072k\pi^4} \[ \gamma - \ln\( \frac{\pi \mu^2}{k^2} \) \] ,
\eea
while the late-time limit, with all time differences $T = t-t_0$, $t'-t_0$, and $t-t'$ satisfying $kT \gg 1$ and $\mu T \gg 1$, exhibits secular growth,
\bea
    \lim_{t \gg t' \gg t_0} G_{k,{\rm 2-loop}}^{-+}(t,t') & = & -\frac{\lambda^2e^{ik(t-t')}}{6144\pi^4k} \ln\left\{ \mu(t-t') \right\} .
\eea
This again matches what we find in \cite{Chaykov:2022zro} and, as shown there, the result can be resummed into a polynomial decay.

Lastly, let us consider the UV-divergent piece of $G_{k,\lambda\phi^4,{\rm ss}}^{-+}(t,t')$ that is given by
\bea
    G_{k,\lambda\phi^4,{\rm ss},\epsilon}^{-+}(t,t') & = & -\frac{\lambda^2 e^{-ik(t-t')}}{6144 \pi^4 k \epsilon} \[ e^{2ik(t-t')} + 3 \] .
\label{eq:Gkphi4epsilon}
\eea
This is again fully canceled by a choice of counterterms similar to that in eq.\ (\ref{eq:ctphi36d}),
\bea
    \delta_m \ = \ 0 \, , \quad \delta_r \ = \ -\frac{\lambda^2}{1536 \pi^4 \epsilon} \, , \quad \delta A_k \ = \ \frac{i\lambda^2k}{1536 \pi^4 \epsilon} \, , \quad \delta B_k \ = \ -\frac{\lambda^2 k}{1536 \pi^4 \epsilon} \, ,
\label{eq:ctphi44d}
\eea
with non-vanishing counterterms in the initial state.

%%%%%%%%%%%%%%%%%%%%%%%%%%%%%%%%%%%%%%%%%%%%%%%%%%

%------------------------------------------------
\section{Discussion}
\label{sec:disc}
%------------------------------------------------

Loop corrections in quantum field theories away from equilibrium exhibit UV divergences beyond those that can be absorbed in standard counterterms in the dynamics. In this paper, we were interested in loop corrections to the unequal-time two-point correlator in different massless self-interacting scalar quantum field theories on a Minkowski background, taking particular care of the counterterms that are needed to cancel all UV divergences. We found finite-time perturbative results using the techniques of in-in perturbation theory, starting the evolution at a finite initial time $t_0$ in the free theory's ground state and allowing for Gaussian corrections to the initial state due to the interaction.

We first considered a $\lambda\phi^3$ interaction in 4D, where $\lambda$ has the dimensions of mass, and found that standard counterterms in the dynamics were sufficient to cancel the UV divergences. We next considered two interactions where $\lambda$ is dimensionless, namely a $\lambda\phi^3$ interaction in 6D and a $\lambda\phi^4$ interaction in 4D. In both cases, we needed additional counterterms in the initial state to cancel all UV divergences. Such additional counterterms are expected to arise due to initial-time singularities that are associated with turning on an interaction at a finite initial time \cite{Calzetta:2008,Baacke:1997zz,Baacke:1999ia,Collins:2005nu,Collins:2014qna}. It is interesting to note, however, that we needed additional counterterms only for the two interactions that have a dimensionless $\lambda$ and are thus marginal. This may indicate a higher degree of entanglement between low and high energy modes for such interactions, along the lines of the results in \cite{Balasubramanian:2011wt}. Interpreting renormalization as tracing out high energy modes would then suggest that the initial state should receive corrections that make it mixed, which is what we find with a non-zero $\delta B$.

For all interactions, we also found that perturbative corrections to the two-point correlator diverge in the late-time limit. In the case of $\lambda\phi^3$ in 4D, we found that the result grows linearly in time, while in the cases of $\lambda\phi^3$ in 6D and $\lambda\phi^4$ in 4D, we found that it grows logarithmically in time. As shown in our companion paper \cite{Chaykov:2022zro}, the WW resummation method exponentiates the late-time result, so that the two-point correlator instead decays exponentially and polynomially, respectively. Lastly, we note again that we restricted to the massless limit as this allowed for analytical calculations. We further left our results in terms of an arbitrary renormalization parameter $\mu$ since we were primarily interested in the UV renormalization and late-time behavior of correlation functions in this paper.

%%%%%%%%%%%%%%%%%%%%%%%%%%%%%%%%%%%%%%%%%%%%%%%%%%

\acknowledgments

We especially thank Daniel Boyanovsky for many insightful discussions and comments on an earlier version of this paper. We also thank Brenden Bowen, Yi-Zen Chu, Mark Hertzberg, and Lorenzo Sorbo for useful conversations. N.~A. and S.~C. were supported by the Department of Energy under award DE-SC0019515. S.~B. was supported by the National Science Foundation under award PHY-1505411, the Eberly research funds of Penn State, and the Urania E. Stott Fund of The Pittsburgh Foundation.

%%%%%%%%%%%%%%%%%%%%%%%%%%%%%%%%%%%%%%%%%%%%%%%%%%

\bibliography{references}

\providecommand{\href}[2]{#2}\begingroup\raggedright\begin{thebibliography}{10}

\bibitem{Schwinger:1960qe}
J.S.~Schwinger, \emph{{Brownian motion of a quantum oscillator}},
  {\emph{J.Math.Phys.} {\bfseries 2} (1961) 407}.

\bibitem{Mahanthappa:1962ex}
K.T.~Mahanthappa, \emph{{Multiple production of photons in quantum
  electrodynamics}},
  \href{https://doi.org/10.1103/PhysRev.126.329}{\emph{Phys.Rev.} {\bfseries
  126} (1962) 329}.

\bibitem{Bakshi:1962dv}
P.M.~Bakshi and K.T.~Mahanthappa, \emph{{Expectation value formalism in quantum
  field theory. 1.}}, {\emph{J.Math.Phys.} {\bfseries 4} (1963) 1}.

\bibitem{Kadanoff:1962}
L.P.~Kadanoff and G.~Baym, \emph{{Quantum Statistical Mechanics}},
  \textnormal{W. A. Benjamin, Inc., New York} (1962).

\bibitem{Bakshi:1963bn}
P.M.~Bakshi and K.T.~Mahanthappa, \emph{{Expectation value formalism in quantum
  field theory. 2.}}, {\emph{J.Math.Phys.} {\bfseries 4} (1963) 12}.

\bibitem{Keldysh:1964ud}
L.V.~Keldysh, \emph{{Diagram technique for nonequilibrium processes}},
  {\emph{Zh.Eksp.Teor.Fiz.} {\bfseries 47} (1964) 1515}.

\bibitem{Jordan:1986ug}
R.D.~Jordan, \emph{{Effective field equations for expectation values}},
  \href{https://doi.org/10.1103/PhysRevD.33.444}{\emph{Phys.Rev.} {\bfseries
  D33} (1986) 444}.

\bibitem{Calzetta:1986ey}
E.~Calzetta and B.L.~Hu, \emph{{Closed time path functional formalism in curved
  space-time: application to cosmological back reaction problems}},
  \href{https://doi.org/10.1103/PhysRevD.35.495}{\emph{Phys.Rev.} {\bfseries
  D35} (1987) 495}.

\bibitem{Calzetta:2008}
E.A.~Calzetta and B.-L.B.~Hu, \emph{{Nonequilibrium quantum field theory}},
  \textnormal{Cambridge University Press} (2008).

\bibitem{Baacke:1997zz}
J.~Baacke, K.~Heitmann and C.~Patzold, \emph{{On the choice of initial states
  in nonequilibrium dynamics}},
  \href{https://doi.org/10.1103/PhysRevD.57.6398}{\emph{Phys. Rev. D}
  {\bfseries 57} (1998) 6398}
  [\href{https://arxiv.org/abs/hep-th/9711144}{{\ttfamily hep-th/9711144}}].

\bibitem{Baacke:1999ia}
J.~Baacke, D.~Boyanovsky and H.J.~de~Vega, \emph{{Initial time singularities in
  nonequilibrium evolution of condensates and their resolution in the
  linearized approximation}},
  \href{https://doi.org/10.1103/PhysRevD.63.045023}{\emph{Phys. Rev. D}
  {\bfseries 63} (2001) 045023}
  [\href{https://arxiv.org/abs/hep-ph/9907337}{{\ttfamily hep-ph/9907337}}].

\bibitem{Collins:2005nu}
H.~Collins and R.~Holman, \emph{{Renormalization of initial conditions and the
  trans-Planckian problem of inflation}},
  \href{https://doi.org/10.1103/PhysRevD.71.085009}{\emph{Phys. Rev. D}
  {\bfseries 71} (2005) 085009}
  [\href{https://arxiv.org/abs/hep-th/0501158}{{\ttfamily hep-th/0501158}}].

\bibitem{Collins:2014qna}
H.~Collins, R.~Holman and T.~Vardanyan, \emph{{Renormalizing an initial
  state}}, \href{https://doi.org/10.1007/JHEP10(2014)124}{\emph{JHEP}
  {\bfseries 10} (2014) 124} [\href{https://arxiv.org/abs/1408.4801}{{\ttfamily
  1408.4801}}].

\bibitem{Chaykov:2022zro}
S.~Chaykov, N.~Agarwal, S.~Bahrami and R.~Holman, \emph{{Loop corrections in
  Minkowski spacetime away from equilibrium 1: Late-time resummations}},
  \href{https://arxiv.org/abs/2206.11288}{{\ttfamily 2206.11288}}.

\bibitem{Berges:2004yj}
J.~Berges, \emph{{Introduction to nonequilibrium quantum field theory}},
  \href{https://doi.org/10.1063/1.1843591}{\emph{AIP Conf. Proc.} {\bfseries
  739} (2005) 3} [\href{https://arxiv.org/abs/hep-ph/0409233}{{\ttfamily
  hep-ph/0409233}}].

\bibitem{Kamenev:2011}
A.~Kamenev, \emph{{Field theory of non-equilibrium systems}},
  \textnormal{Cambridge University Press} (2011).

\bibitem{Weinberg:2005vy}
S.~Weinberg, \emph{{Quantum contributions to cosmological correlations}},
  \href{https://doi.org/10.1103/PhysRevD.72.043514}{\emph{Phys. Rev.}
  {\bfseries D72} (2005) 043514}
  [\href{https://arxiv.org/abs/hep-th/0506236}{{\ttfamily hep-th/0506236}}].

\bibitem{Agarwal:20xx}
N.~Agarwal and Y.-Z.~Chu, \emph{{Initial value formulation of a quantum damped
  harmonic oscillator}}, \textnormal{In preparation}.

\bibitem{Agarwal:2012mq}
N.~Agarwal, R.~Holman, A.J.~Tolley and J.~Lin, \emph{{Effective field theory
  and non-Gaussianity from general inflationary states}},
  \href{https://doi.org/10.1007/JHEP05(2013)085}{\emph{JHEP} {\bfseries 1305}
  (2013) 085} [\href{https://arxiv.org/abs/1212.1172}{{\ttfamily 1212.1172}}].

\bibitem{Chen:2016nrs}
X.~Chen, Y.~Wang and Z.-Z.~Xianyu, \emph{{Loop corrections to standard model
  fields in inflation}},
  \href{https://doi.org/10.1007/JHEP08(2016)051}{\emph{JHEP} {\bfseries 08}
  (2016) 051} [\href{https://arxiv.org/abs/1604.07841}{{\ttfamily
  1604.07841}}].

\bibitem{Balasubramanian:2011wt}
V.~Balasubramanian, M.B.~McDermott and M.~Van~Raamsdonk, \emph{{Momentum-space
  entanglement and renormalization in quantum field theory}},
  \href{https://doi.org/10.1103/PhysRevD.86.045014}{\emph{Phys. Rev. D}
  {\bfseries 86} (2012) 045014}
  [\href{https://arxiv.org/abs/1108.3568}{{\ttfamily 1108.3568}}].

\end{thebibliography}\endgroup



\providecommand{\href}[2]{#2}\begingroup\raggedright\endgroup
\bibliographystyle{JHEP}

\end{document}